\documentclass[aps,prl,reprint,groupedaddress]{revtex4-2}

\usepackage{graphicx}
\usepackage{subfigure}
\usepackage{float}
\usepackage{dcolumn}
\usepackage{bm}
\usepackage{caption}

\begin{document}


\title{Single-Shot Laser-Driven Neutron Resonance Spectroscopy for Temperature Profiling}


\author{Zechen Lan, Yasunobu Arikawa, S. Reza Mirfayzi$^{2}$, Alessio Morace, Takehito Hayakawa$^{3}$, Hirotaka Sato$^{4}$, Takashi Kamiyama$^{4}$, Tianyun Wei, Yuta Tatsumi, Mitsuo Koizumi$^{5}$, Yuki Abe$^{6}$, Shinsuke Fujioka, Kunioki Mima, Ryosuke Kodama and Akifumi Yogo}
\email[]{yogo.akifumi.ile@osaka-u.ac.jp}
\affiliation{$^{1}$Institute of Laser Engineering, Osaka University, Japan}
\affiliation{$^{2}$Tokamak Energy Ltd, UK}
\affiliation{$^{3}$National Institute for Quantum Science and Technology, Japan}
\affiliation{$^{4}$Faculty of Engineering, Hokkaido University, Japan}
\affiliation{$^{5}$Japan Atomic Energy Agency, Japan}
\affiliation{$^{6}$Institute of Laser Engineering, Osaka University, Japan}

\date{\today}

\begin{abstract}
The temperature measurement of material inside of an object is one of the key technologies for control of dynamical processes. For this purpose, various techniques such as laser-based thermography \cite{yoneyama2018three} and phase-contrast imaging thermography \cite{ahmadi2020laser} have been studied. However, it is, in principle, impossible to measure the temperature of an element inside of an object using these techniques. One of the possible solutions is measurements of Doppler brooding effect in neutron resonance absorption (NRA). Here we present a method to measure the temperature of an element or an isotope inside of an object using NRA with a single neutron pulse of approximately 100 ns width provided from a high-power laser. We demonstrate temperature measurements of a tantalum (Ta) metallic foil heated from the room temperature up to 617 K. Although the neutron energy resolution is fluctuated from shot to shot, we obtain exactly the temperature using a reference of a silver (Ag) foil kept to the room temperature. A free gas model well reproduces the results. This method enables element(isotope)-sensitive thermometry to detect the instantaneous temperature rise in dynamical processes.
\end{abstract}


\maketitle

\section{Introduction}
The NRA is the process in which a neutron at an energy related to the excited states of the atomic nucleus just above the neutron binding energy is resonantly captured by the nucleus. Thus, the probability that neutrons interact with nuclei depends on the incident energy and atom identity \cite{breit1936capture}. 
Bethe \cite{bethe1937nuclear} suggested the temperature dependence of the broadening of a resonance peak due to the Doppler effect. Thus, the neutron resonance spectroscopy (NRA) provides a possibility of detecting the temperature as well as the elemental composition and density inside a material non-destructively. 

The NRS has been predominantly implemented with accelerator-based neutron sources for material analysis \cite{hasemi2014quantitative,schillebeeckx2012neutron,tremsin2017non}, spatially resolved thermometry \cite{tremsin2015spatially,kamiyama2005computer,hasemi2014quantitative} and shock wave measurement \cite{yuan2005shock,ragan1977shock}. In these methods, neutron energy spectra are measured by using the time-of-flight (TOF) method with a pulsed neutron beam. A pulsed neutron source is required for the TOF to ensure all the neutrons start at same time. Nevertheless, it's impractical to generate a neutron pulse with perfect simultaneity, resulting in a finite pulse duration that introduces measurement errors in the determination of flight time. To obtain sufficient energy resolution, a long beamline which is typically a few tens of metres for accelerator-based neutron sources \cite{anderson2016research}, must be set for TOF measurements. However, a long beamline reduces the flux of neutrons arriving at the detector due to the spherically emission of pulsed neutron sources. To obtain a neutron spectrum with sufficient statistics, neutron bunches must be integrated for several hours in typical neutron facilities. As a result, even if the pulse width and flux of accelerator-based neutron sources are stable, the energy resolution becomes lower than that of a single neutron pulse.

As an alternative sources, laser-driven neutron sources (LDNSs) \cite{willingale2011comparison,higginson2011production,roth2013bright,jung2013characterization,pomerantz2014ultrashort,kar2016beamed,alejo2017high,kleinschmidt2018intense,gunther2022forward,yogo2023laser} have attracted widespread attention for their compactness and short pulse performance. An LDNS utilizes laser-driven ion acceleration \cite{clark2000energetic,maksimchuk2000forward,snavely2000intense,passoni2010target,yogo2017boosting}, which enables the acceleration of protons or deuterons up to a few tens of MeV. By placing a secondary target, typically beryllium (Be) or lithium (Li), in the beam path, a pulse of MeV-energy neutrons is generated by nuclear reactions between the target materials and laser-accelerated ions. A pulsed fast neutron beam can be generated by the LDNS of$\sim$ 4 cm size within a pulse duration of $\sim$ 1 ns. A currently highest neutron yield of $\sim$ $10^{11}$ has been achieved \cite{roth2013bright,alejo2017high,kleinschmidt2018intense,yogo2023laser} for a single pulse. Furthermore, LDNSs generated low-energy neutrons at eV $\sim$ meV region have been generated by employing neutron moderators made of hydrogen-rich materials at room temperature \cite{higginson2010laser,mirfayzi2020miniature,zimmer2022demonstration,yogo2023laser} or cryogenically cooled solid hydrogen \cite{mirfayzi2020proof}. The short pulse duration and miniature scale of an LDNS allow a smaller neutron moderator, resulting in a reduction of the time during which the neutron pulse expands in the moderation process. At present, accelerator-based neutron sources are more stable than LDNSs. However, when a single laser shot is used for measuring the NRA, the energy resolution originating from the short-term expansion of the initial neutron pulse is expected to be higher than that using an accelerator-based neutron source.
 
Previous studies demonstrated NRA using LDNS \cite{kishon2019laser,zimmer2022demonstration,mirani2023laser}. In our last work \cite{yogo2023laser}, we measured neutron resonances in the epithermal (several eV) region using a 1.8 m beamline after an LDNS, where one resonance spectrum was obtained with a single bunch of neutrons generated by a single pulse of the laser.

In this study, we report measurements of the Doppler broadening of a neutron resonance absorption generated with a single neutron pulse provided by laser. We demonstrate that the Doppler width of the resonance absorption of an atomic nucleus $^{181}$Ta increases as the square root of the sample temperature according to a theoretical model. Each Doppler spectrum is acquired from a single pulse of a laser. This result indicates the possibility that an LDNS may provide a real-time thermometry of a nuclide that probes the instantaneous temperature of dynamic objects. 

\section{Neutron Generation and Moderation}

\begin{figure}[htbp]
	\begin{center}
	\includegraphics[width=0.5\textwidth]{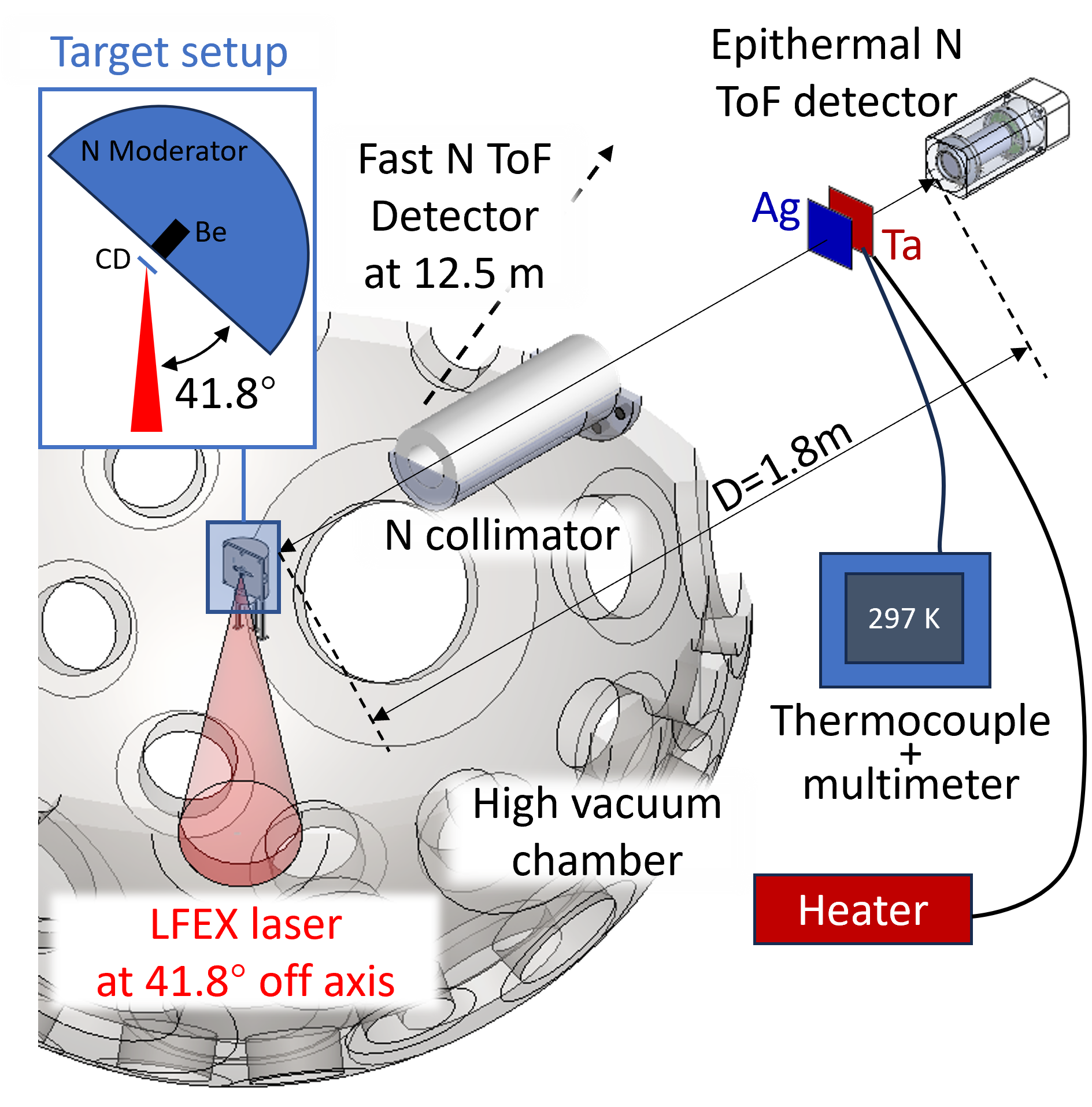}%
	\end{center}
	\caption{\label{Fig.1} The experimental setup of the laser-driven epithermal neutron generation and resonance absorption measurement using the ToF method.}
\end{figure}

We demonstrate a proof-of-principle experiment for the neuron thermometry with the laser-driven NRA. The experiment was carried out using 1.5 ps laser pulses provided from Fast Ignition Experiment (LFEX) laser system \cite{miyanaga200610}, delivering a total energy of 900 J on the target at the Institute of Laser Engineering (ILE), Osaka University. The setup [Fig. \ref{Fig.1}] includes a vacuum chamber. A laser pulse with an intensity of $\sim$ 1 $\times$ $10^{19}$  W$\cdot$cm$^{-2}$ was focused on a 5 $\mu$m thick deuterated polystyrene (CD) foil to accelerate protons and deuterons \cite{yogo2023laser}. Before NRA measurement, the energies of the ions were measured by a Thomson parabola (TP) spectrometer \cite{golovin2021calibration} located in the normal direction of the CD foil. The protons and deuterons were accelerated up to MeV energy, where the ps duration laser pulse enhanced the acceleration of the deuterons \cite{yogo2023laser}.
To generate neutrons, the ions were injected into a secondary target, a Be cylinder with a diameter of 10 mm and a thickness of 10 mm. The diameter is larger than the angular divergence of the incident ions. The thickness is sufficient to stop the ions. Fast neutrons with energy over tens of MeV were generated via the $^{9}$Be(p,n)$^{9}$B and $^{9}$Be(d,n)$^{10}$B nuclear reactions.
The maximum energy of the fast neutrons reaches 25 MeV, with a neutron number of $2.3 \times 10^{10}$ n/sr\cite{yogo2023laser}. The total number of neutrons in the 4$\pi$ direction reaches $10^{11}$ in a single laser shot.

The NRA could be identified in the energy region of eV [Fig. \ref{Fig.4(b)}]. For the temperature changing within 300 - 1000 K, sub-eV Doppler broadening width could be well investigated by NRS in eV energy region. Therefore, we located a neutron moderator on the LDNS to generate the eV neutrons. The moderator was made from high-density polyethylene (0.98 g$\cdot$cm$^{-3}$) in the shape of a cylinder and was attached to the Be target. The neutrons are moderated via elastic collisions with the nuclei of moderator materials. The kinetic energies of neutrons are reduced most efficiently in the collisions with hydrogen nuclei. The moderated neutrons passed through an aluminum window into air and their energies were measured at 1.78-m distance by a $^{6}$Li-TOF neutron detector consisting of a $^{6}$Li-doped glass scintillator (GS20, Scintacor Ltd., 10 mm in thickness) coupled to a time-gated PMT developed based on a HAMAMATSU-R2083. To protect the PMT and ensure the linearity of the output response, the time gate was set to block the electromagnetic pulses and the intense flash of X-rays generated by the laser-plasma interactions. To reduce the background due to neutrons scattered by the chamber wall, we installed an nickel (Ni) collimator on the beamline.

We set a Ag metallic foil and a Ta metallic foil in the beamline to measure NRA of $^{109}$Ag at 5.19 eV and $^{181}$Ta at 4.28 eV. The thickness of Ag and Ta target are 0.2 mm and 0.1 mm, respectively. These materials are well-suited for demonstration purposes because their resonance peaks exhibit a large reaction cross sections and are not affected by nearby resonances. The Ta target was heated to $T = 361, 413, 474, 573$ and $617$ K, which is the highest temperature achievable with the present heating system. The temperature was monitored by a thermocouple, as shown in Fig. \ref{Fig.1}. The Ag sample was kept at room temperature to provide a reference of the neutron spectrum. The NRA were measured for each temperature.

\begin{figure}[htbp]
	\subfigbottomskip=6pt
	\subfigcapskip=0pt
	\begin{center} 
		\subfigure[\label{Fig.3(a)}]{
			\includegraphics[width=0.45\textwidth]{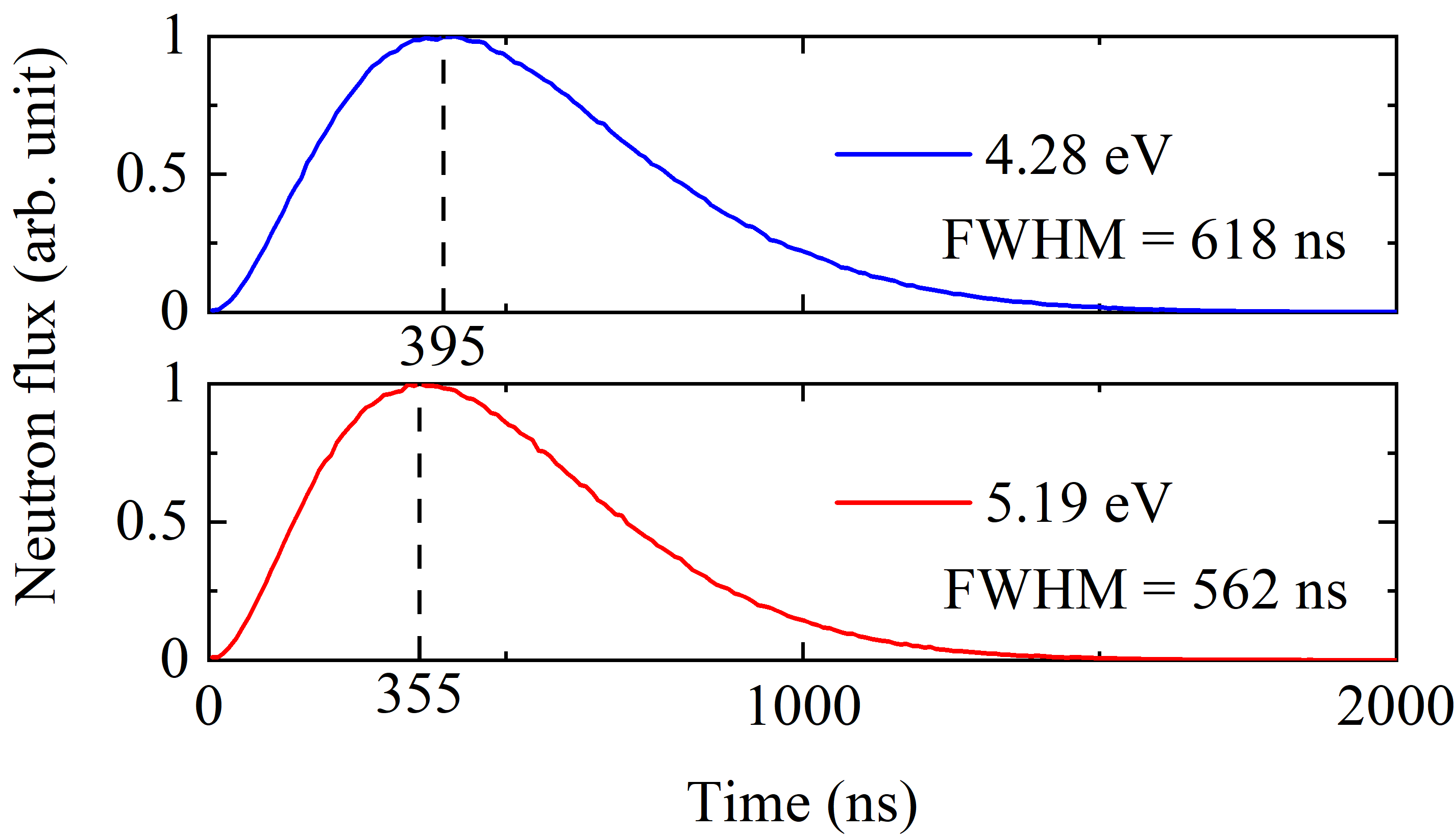}
		}

		\subfigure[\label{Fig.3(b)}]{
			\includegraphics[width=0.45\textwidth]{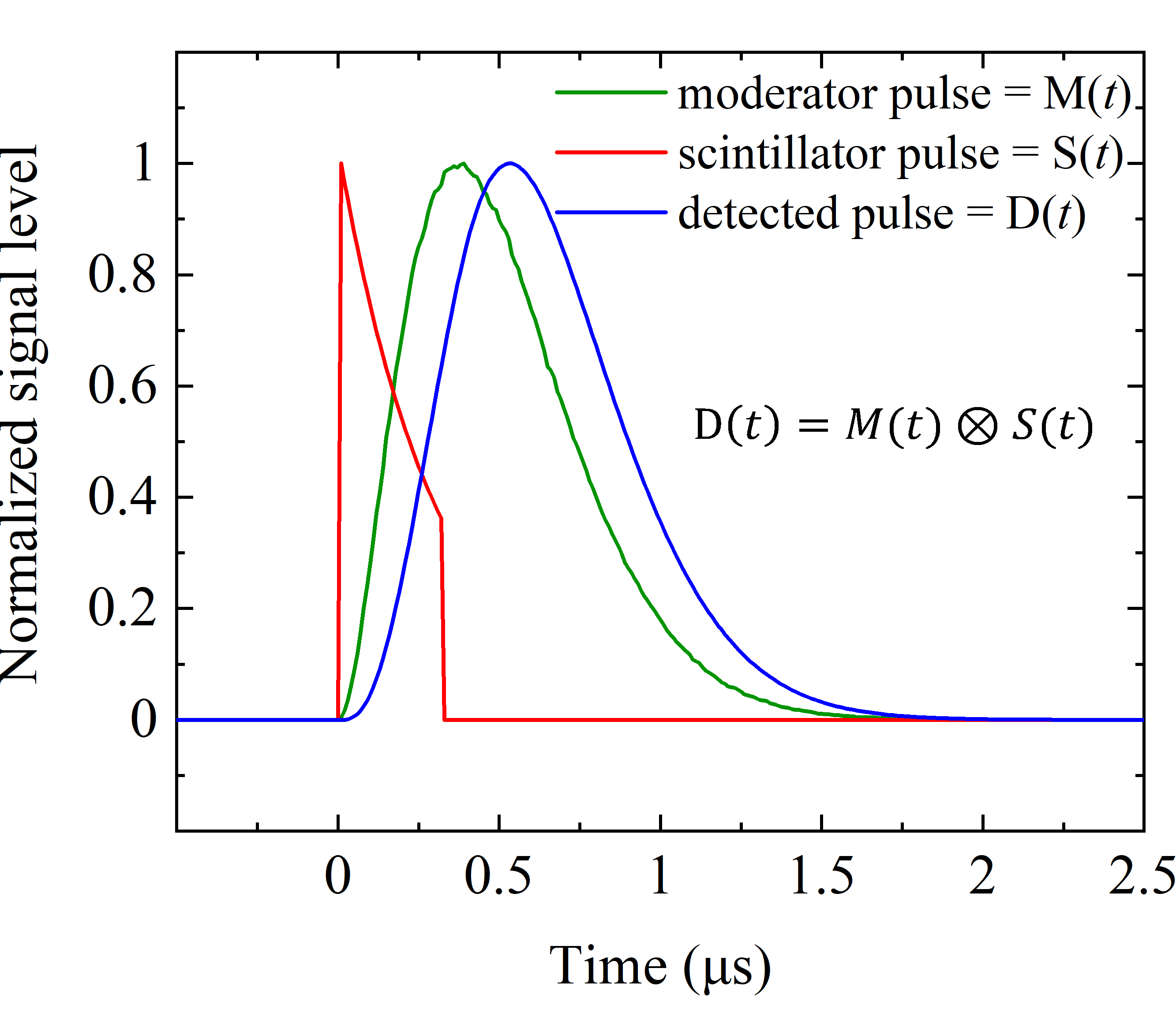}
		}
	\end{center}
	\caption{\label{Fig.3}(a) Simulated pulse duration for 5.19 eV and 4.28 eV neutrons. The vertical dotted lines are marking the peak timing of the neutron pulse. (b) Pulse duration at the simulated neutron moderator (green line), pulse duration at the $^{6}$Li glass scintillator (red line), and convoluted result at the detector at 1.78 m (blue line, systematic error).}
\end{figure}

\section{Single-shot analysis of Neutron Resonance} 

To evaluate the temporal structure of the moderated neutron pulses, we used the PHITS Monte Carlo code \cite{sato2018features}. In the simulation, the neutrons are monitored at the exit of the moderator. The time duration of the moderation process is not negligible relative to the total flight time to the detector. Fig. \ref{Fig.3(a)} shows the temporal structure of moderated neutrons M(t) with kinetic energies 5.19 eV and 4.28 eV, which correspond to the resonance energies of $^{109}$Ag and $^{181}$Ta, respectively. Here, time zero indicates the time when the laser is first incident on the target, and the time for neutron generation in the LDNS is shorter than 50 ps \cite{roth2013bright,pomerantz2014ultrashort,yogo2023laser,jung2013characterization}. The pulse broadening is evaluated to be 562 and 618 ns at the FWHM for 5.19 and 4.28 eV neutrons, respectively, based on the temporal structures [Fig. \ref{Fig.3(a)}]. The pulse broadening at the $^{6}$Li-doped scintillator S(t) is evaluated to be 320 ns for 5 eV neutrons [red line in Fig. \ref{Fig.3(b)}]. By convoluting the temporal broadening induced by the moderator $M(t)$ [green line in Fig. \ref{Fig.3(b)}] and the scintillator S(t), the total pulse broadening effect in the neutron detection is obtained as $D(t) = M(t) \otimes S(t)$ [blue line in Fig. \ref{Fig.3(b)}], where $\otimes$ represents the convolution operation.

By using the D(t) function, we obtain TOF signals with the pulse broadening effects at the moderator and the detector being removed in every laser shot. Fig. \ref{Fig.4(a)} shows the signal of epithermal neutrons transmitted through Ag and Ta plates at the room temperature recorded by the $^{6}$Li-TOF detector. In the time region of 50 - 60 $\mu$s, two distinct dips are seen in the continuous neutron signal. The flight time $t_{raw}$ measured by the detector is expressed by
	$t_{raw} = t \otimes D(t)$, 
where $t$ is the flight time deconvoluved from the pulse broadening $D(t)$. 
The neutron kinetic energy $E$ is obtained by $E = m(d/t)^{2}/2$, where $d = 1.78$ m is the flight distance, and $m$ is the mass of the neutron. The energies of the two dips correspond to the resonance energies of 4.28 eV and 5.19 eV for $^{181}$Ta and $^{109}$Ag, respectively [Fig. \ref{Fig.4(b)}].
The TOF signal is converted into the neutron absorption rate $R_{exp}(E)$ [dashed line in Fig. \ref{Fig.4(c)}], where we assume $R_{exp}(E) = 0$ for the base line [dashed line in Fig. \ref{Fig.4(a)}] \cite{mirfayzi2017experimental}.
The absorption rate $R_{0}(E)$ [red line in Fig. \ref{Fig.4(b)}] could be obtained by
\begin{equation}
	\label{equ.1}
	R_{0}(E) = \sum_{Ag, Ta} \exp(-nl\sigma(E)), 
\end{equation}
where $n$ is the atomic areal number density, $l$ is the thickness of the targets and $\sigma(E)$ is the NRA cross section (JENDL4.0 \cite{shibata2011jendl}) at room temperature.
As a comparison, the black line in Fig. \ref{Fig.4(c)} shows the theoretical absorption rate when $t_{raw} = t$ without considering the effect of $D(t)$. $R_{exp}(E)$ exhibits two peaks at 4.3 and 5.2 eV consistent with the resonance energies for $^{181}$Ta and $^{109}$Ag, respectively, which are also found in the analytical model $R_{0}(E)$. 
However, the detailed shape of the experimental peaks is not well reproduced by $R_{0}(E)$, especially for the thicker target (Ag), where the saturated absorption observed in $R_{0}(E)$ is not exhibited in $R_{exp}(E)$.
This difference indicates the presence of another effect, $F(t)$, that causes further pulse broadening in addition to $D(t)$. 
The pulse broadening caused by $F(t)$. $F(t)$ is considered to originate from statistical neutron scattering in the beamline from the neutron moderator and involved in a Gaussian form.
We determine $F(E)$ as a function of E by fitting the experimental absorption rate $R_{exp}(E)$ with the following equation:
\begin{equation}
	\label{equ.2}
	R_{1}(E) = R_{0}(E) \otimes F(E), 
\end{equation}
Fig. \ref{Fig.6} shows $R_{exp}(E)$ (gray) fitted with the model $R_{1}(E)$ (red) by using an nonlinear-curve-fitting function, where the resonance peaks are well reproduced as the errors shown in the lower frame. $F(t)$ has a half-width of approximately 100 $\sim$ 200 ns, which is sufficiently shorter than $D(t)$ ($\sim$ 0.5 $\mu$s). This result indicates that the advantage of the miniature size of the moderator is not offset by the pulse broadening caused along the beamline.

\section{Temperature Dependence}
\begin{figure}[htbp]
	\subfigbottomskip=0pt
	\subfigcapskip=0pt 
	\begin{center}
		\subfigure[\label{Fig.4(a)}]{
			\includegraphics[width=0.3\textwidth]{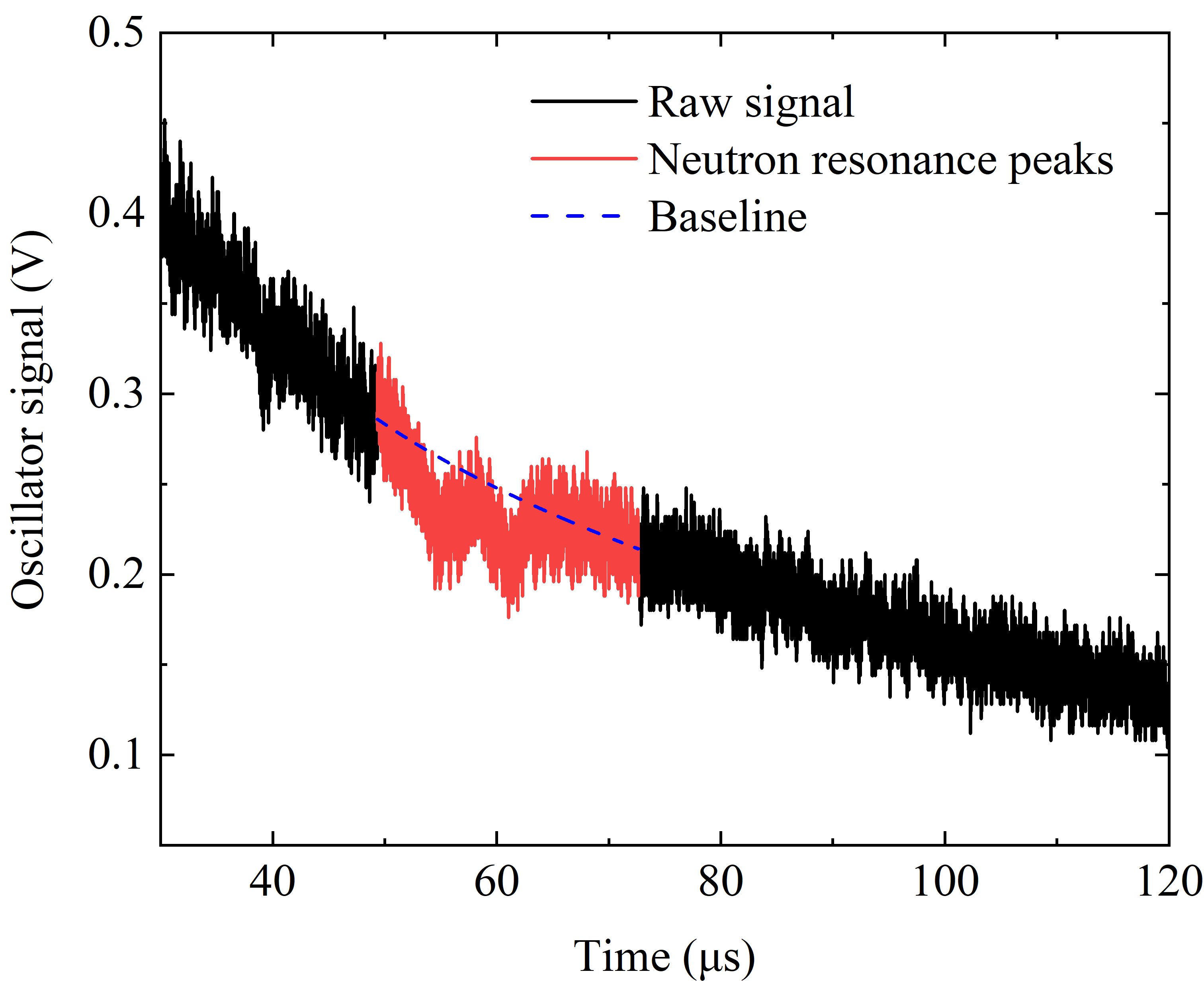}
		}
		\subfigure[\label{Fig.4(b)}]{
			\includegraphics[width=0.3\textwidth]{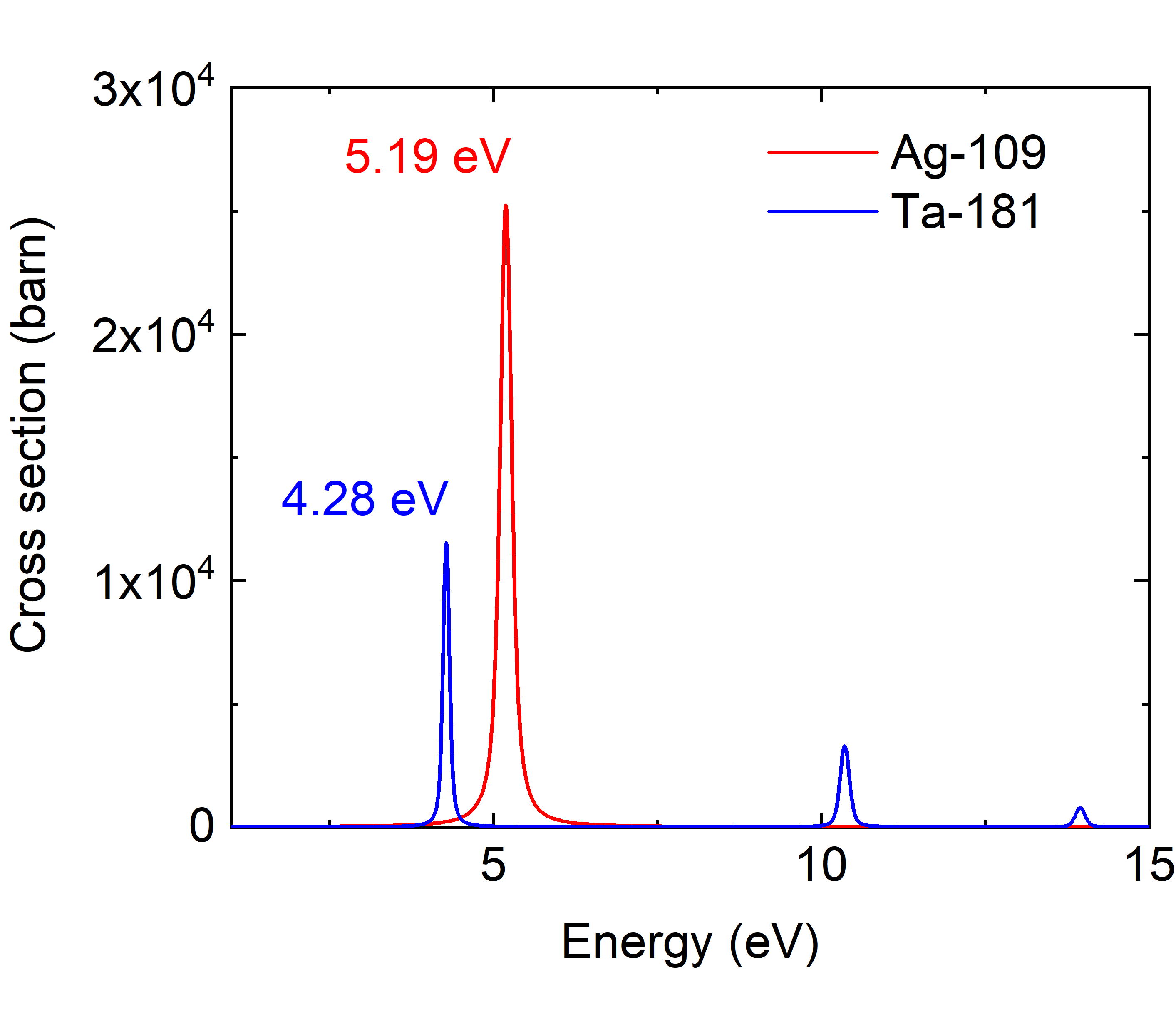}
		}
		\subfigure[\label{Fig.4(c)}]{
			\includegraphics[width=0.3\textwidth]{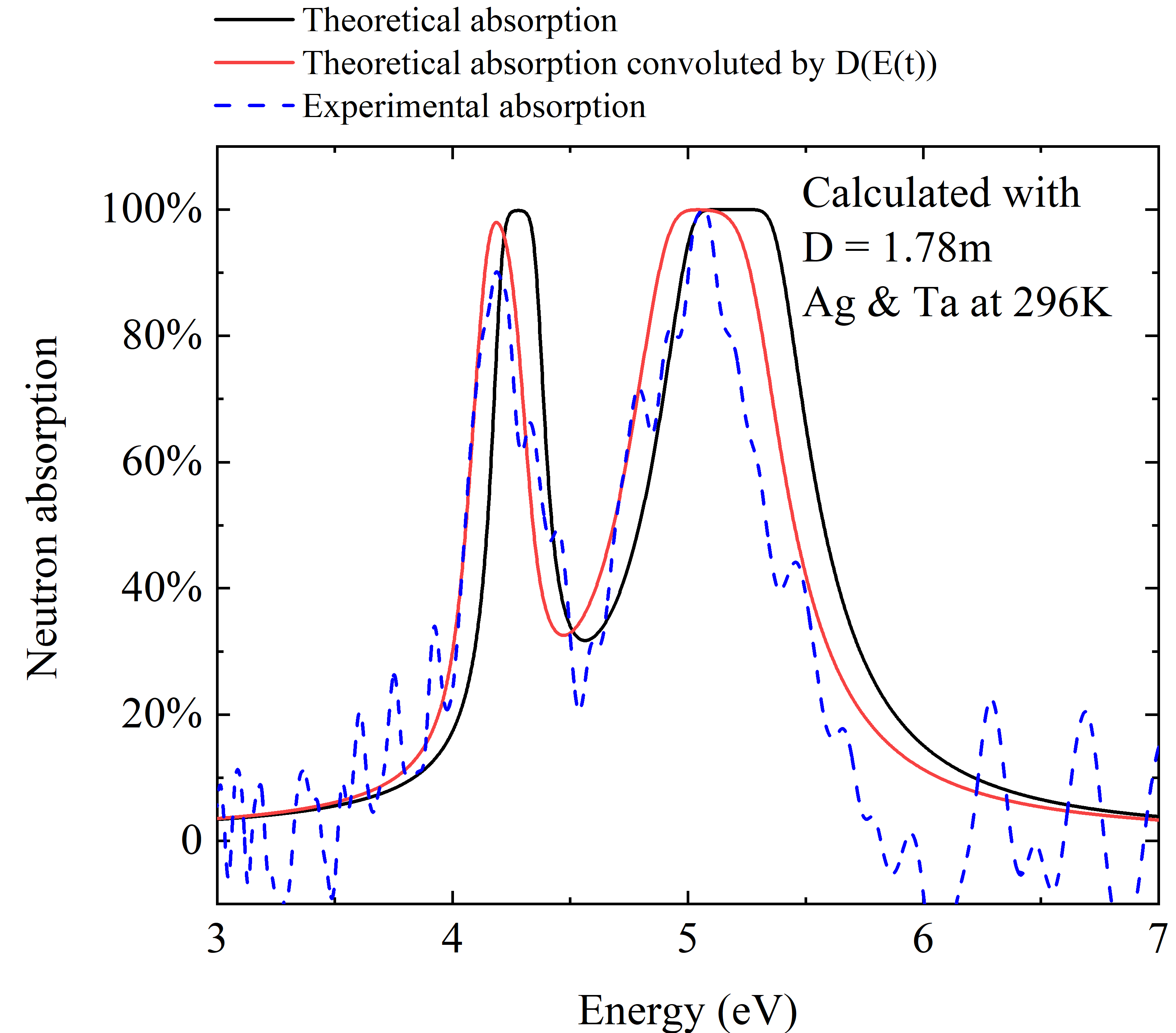}%
		} 
		\subfigure[\label{Fig.4(d)}]{
			\includegraphics[width=0.3\textwidth]{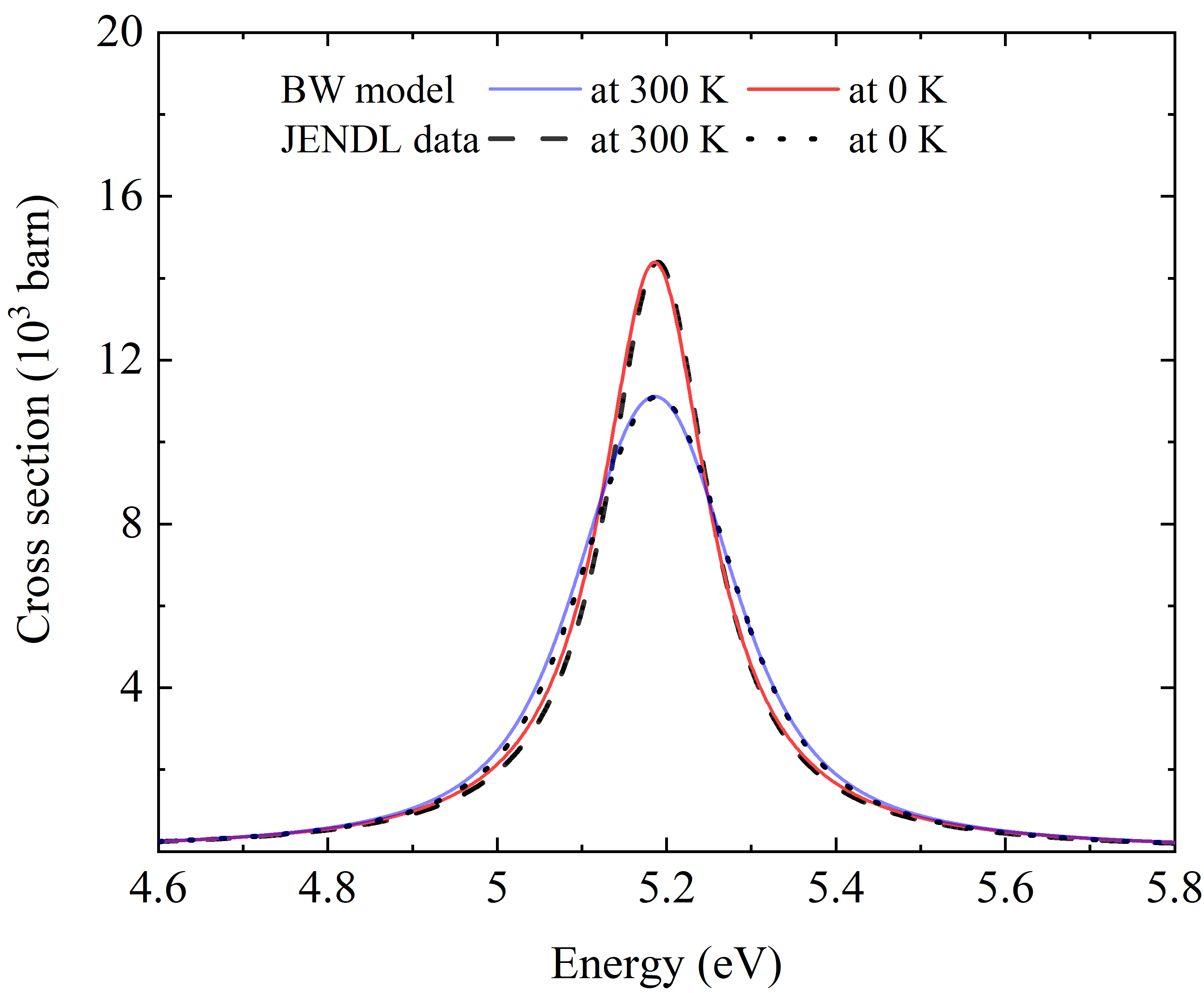}%
		} 
	\end{center}
	\caption{\label{Fig.4}(a) The epithermal neutron signal with both time and energy axes. Two troughs can be checked in a continuous curve which should be exponential. (b) The neutron absorption cross sections of $^{109}$Ag and $^{181}$Ta obtained from JENDL4.0 data base \cite{shibata2011jendl}. (c) Experimental and theoretical neutron absorption with energy. (d) JEDNL4.0 cross section data and $\sigma_{BW}(E)$ calculation of a $^{109}$Ag resonance at 5.19 eV.}

\end{figure}

To analyse the temperature dependence of the resonance absorption, we introduce a Breit-Wigner (BW) single-level formula \cite{breit1936capture}, as seen in \cite{yuan2005shock,tremsin2014neutron}, for the neutron absorption cross section at 0 K:
\begin{equation}
	\label{equ.3}
	\sigma_{BW}(E^{\prime}) = \pi\lambdabar \mathit{g_{j}}\frac{\Gamma_{n} \Gamma_{\gamma}}{(E^{\prime}-E_{r})^{2}+(\Gamma_{n} + \Gamma_{\gamma})^{2}/4},
\end{equation}
where $E_{r}$ is the resonance energy, $E^{\prime}$ is the kinetic energy of the incident neutron relative to the target nucleus with $E^{\prime} = E$ at 0 K, $\lambdabar$ is the de Broglie wavelength of incident neutron (divided by $2\pi$), and $\mathit{g_{j}}$ is a statistical factor determined by the angular momentum. $\Gamma_{n}$ and $\Gamma_{\gamma}$ represent the resonance width for the neutron and decay width for $\gamma$-ray, respectively.
For the $^{181}$Ta resonance at $E_{r}=4.28$ eV, $\Gamma_n=1.74$ meV and $\Gamma_{\gamma}=55$ meV; For the $^{109}$Ag resonance at $E_(r)$ = 5.19 eV, $\Gamma_{n}$ = 8.34 meV, and $\Gamma_{\gamma}$ = 136 meV \cite{shibata2011jendl}. We have confirmed the $\sigma_{BW}(E)$ formula by fitting it to the latest analytical result of JENDL4.0 data base \cite{shibata2011jendl} for 0 K and 300 K [Fig. \ref{Fig.4(d)}].

When the thermal motion of target nuclei is sufficiently slower than the incident neutron, the Doppler broadening effect can be well approximated by a Gaussian function \cite{yuan2005shock,tremsin2015spatially}.
We develop an analytical cross-section $\sigma_{T}(E)$ for the neutron resonance absorption involving the target temperature as follows
\begin{equation}
	\label{equ.5}
	\sigma_{T}(E) = \sigma_{BW}(E) \otimes A \exp[-\frac{(E-E_{r})^{2}}{2\Gamma_{D}(T)^{2}}].
\end{equation}
where $A$ is a fitting parameter, and the width $\Gamma_{D}(T)$ broadens as the temperature $T$. 
Then, a temperature-dependent absorption rate $R_{T}(E)$ is developed as follows:
\begin{equation}
	\label{equ.6}
	R_{T}(E) = \sum_{Ag, Ta} \exp(-nl\sigma_{T}(E)) \otimes F(E).
\end{equation}

\begin{figure*}[htbp]
	\subfigbottomskip=6pt
	\subfigcapskip=0pt 
	\begin{center}
		\subfigure[\label{Fig.6(a)}]{
		\includegraphics[width=0.8\textwidth]{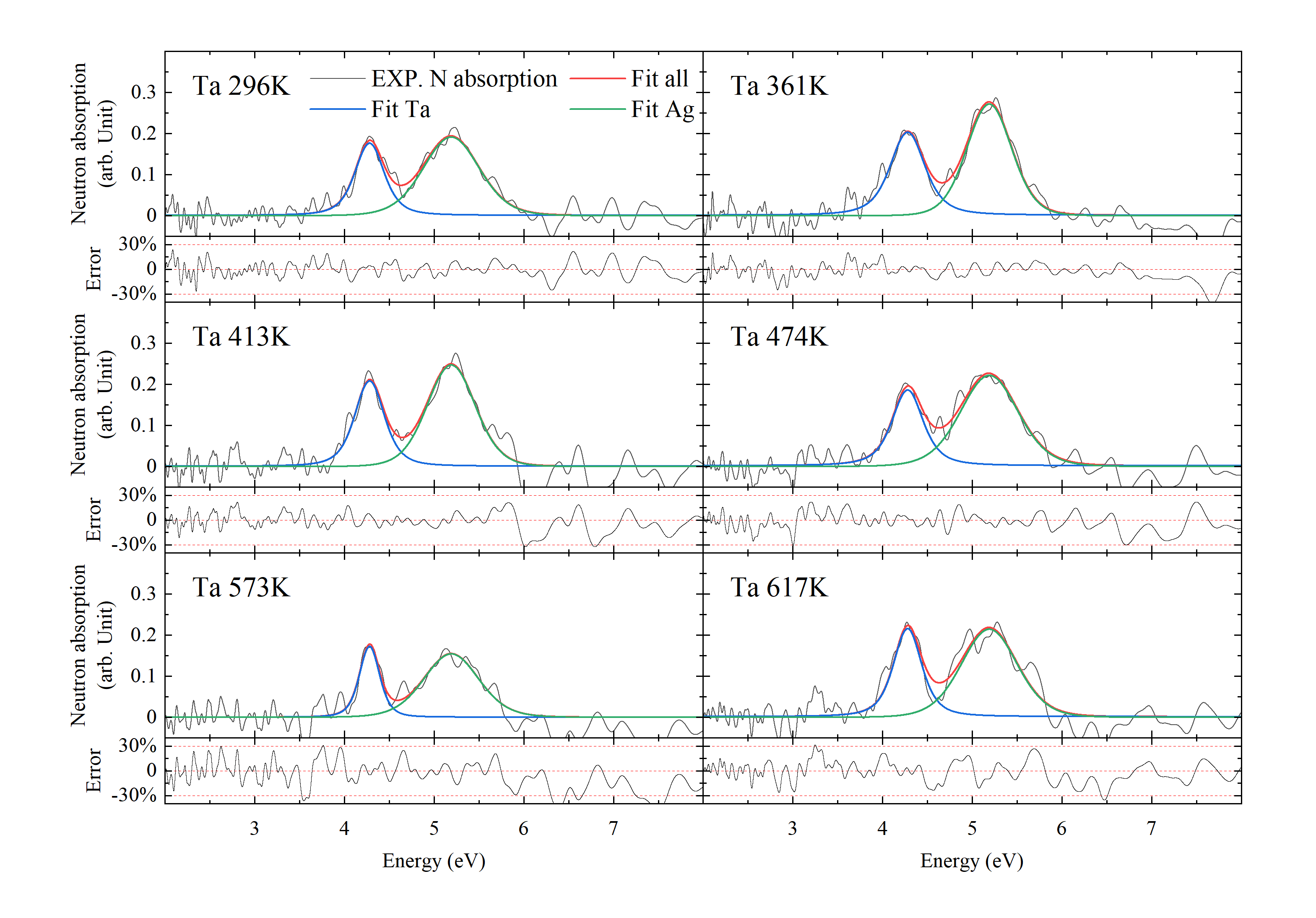}
		}
		\subfigure[\label{Fig.6(b)}]{
			\includegraphics[width=0.5\textwidth]{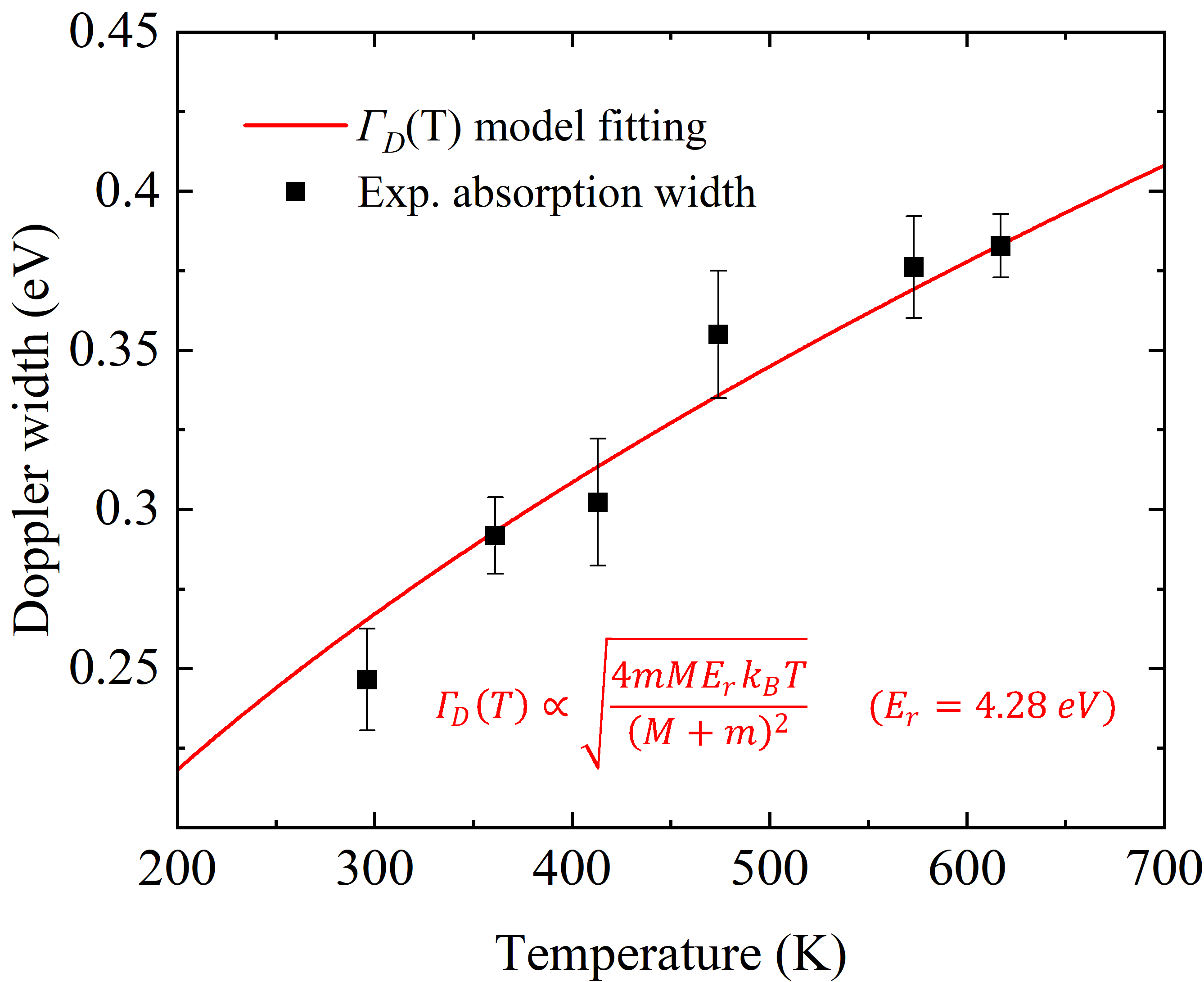}
		}
		\caption{\label{Fig.6} (a) Experimental neutron absorption results and model fitting by $R_{T}(E)$. The temperature of Ag was kept at 296K and Ta was heated to $T = 297, 361, 413, 474, 573$ and $617$ K. (b) Theoretical Doppler broadening width and experimental results. The error bars are depended by fitting error and noise level of original signal. A tolerance area is defined by the errors, showing a estimated Doppler width range with experimental temperature.}
	\end{center}
\end{figure*}

According to the model given by Eq.(\ref{equ.5}), we analyze the absorption rate $R_{exp}(E)$ measured for different target temperature $T$, as shown in Fig. \ref{Fig.6(a)}.
Although the Ag target was kept at room temperature, it can be seen that the resonance width of $^{109}$Ag varies from shot to shot. This is due to the fluctuation of $F(E)$ caused by statistical processes including neutron scattering on the beamline. Therefore, the $^{109}$Ag resonance serves as a reference to evaluate $F(E)$ for each measurement. In Fig. \ref{Fig.6(b)}, we plot the Doppler width $\Gamma_{D}(T)$ obtained by fitting $R_{exp}(E)$ as a function of $T$. The error bars are determined by the least mean square of the difference between the fitted curves and experimental data.
The Doppler width increases as the square root of $T$. We also show a calculated result using a free gas model by Bethe \cite{bethe1937nuclear}:
\begin{equation}
	\label{equ.4}
\Gamma_{D}(T) \propto \sqrt{\frac{2mM}{(M+m)^{2}} E_{r} k_{B} T},
\end{equation}
where $m$ and $M$ are the masses of the neutron and the target nucleus, respectively, and $k_{B}$ is Boltzmann's constant. This model can well reproduce the experimental results.

\section{Conclusion}
The temperature dependence of NRA is experimentally demonstrated by using single shot of our LDNS. This result shows that the temperature of an element or an isotope inside of a material can be measured using NRA with a single neutron pulse of approximately 100 ns provided from a LDNS. The shortness of the moderated neutron pulse contributes to shortening of the neutron beamline with the high energy resolution in the TOF method, which enhances the number of neutrons arriving at a detector. Pumping laser systems operating at 10–100 Hz \cite{divoky2021150,ogino202110} will further improve
the repetition rate, which indicates that \textit{real-time} isotope-sensitive thermometry at a same frequency of laser.
\section{Acknowledgment}

This work was funded by JSPS KAKENHI Grant-in-Aid for Scientific Research (JP25420911, JP26246043, JP22H02007, JP22H01239), JST A-STEP (AS2721002c) and JST PRESTO (JPMJPR15PD).
Zechen Lan was supported by JSPS Research Fellowship for Young Scientists DC2 (202311207). Zechen Lan (before Apr. 2023) and Tianyun Wei were supported by JST SPRING, Grant Number JPMJSP2138.
The authors thank the technical support staff of ILE for their assistance with the laser operation, target fabrication and plasma diagnostics. This work was supported by the Collaboration Research Program of ILE, Osaka University.

\bibliography{ResEXP2020bib_TH}

\end{document}